\input harvmac
\noblackbox

\font\cmss=cmss10 \font\cmsss=cmss10 at 7pt
 \def\inbar{\,\vrule height1.5ex width.4pt depth0pt}
\def\IZ{\relax\ifmmode\mathchoice
{\hbox{\cmss Z\kern-.4em Z}}{\hbox{\cmss Z\kern-.4em Z}}
{\lower.9pt\hbox{\cmsss Z\kern-.4em Z}}
{\lower1.2pt\hbox{\cmsss Z\kern-.4em Z}}\else{\cmss Z\kern-.4em
Z}\fi}
\def\IB{\relax{\rm I\kern-.18em B}}
\def\IC{{\relax\hbox{$\inbar\kern-.3em{\rm C}$}}}
\def\ID{\relax{\rm I\kern-.18em D}}
\def\IE{\relax{\rm I\kern-.18em E}}
\def\IF{\relax{\rm I\kern-.18em F}}
\def\IG{\relax\hbox{$\inbar\kern-.3em{\rm G}$}}
\def\IGa{\relax\hbox{${\rm I}\kern-.18em\Gamma$}}
\def\IH{\relax{\rm I\kern-.18em H}}
\def\II{\relax{\rm I\kern-.18em I}}
\def\IK{\relax{\rm I\kern-.18em K}}
\def\IC{\relax{\rm I\kern-.18em C}}
\def\IR{\relax{\rm I\kern-.18em R}}


\Title{\vbox{\baselineskip12pt\hbox{hep-th/9810129}
\hbox{LBNL-42248, SLAC-PUB-7970, UCB-PTH-98/46}
}}
{\vbox{\centerline{
On Vanishing Two Loop Cosmological Constants}\smallskip
\centerline{in Nonsupersymmetric Strings}}
}
\centerline{Shamit Kachru$^{1}$ and Eva Silverstein$^{2}$ 
}
\bigskip
\bigskip
\centerline{$^{1}$Department of Physics}
\centerline{University of California at Berkeley}
\centerline{Berkeley, CA 94720}
\smallskip
\centerline{and}
\smallskip
\centerline{Ernest Orlando Lawrence Berkeley National Laboratory}
\centerline{Mail Stop 50A-5101, Berkeley, CA 94720}
\medskip
\medskip
\centerline{$^{2}$ Stanford Linear Accelerator Center}
\centerline{Stanford University}
\centerline{Stanford, CA 94309}
\bigskip
\medskip
\noindent

It has recently been suggested that in certain special nonsupersymmetric
type II string compactifications, at least 
the first two perturbative contributions
to the cosmological constant $\Lambda$ 
vanish.  Support for perturbative vanishing
beyond 1-loop 
(as well as evidence for the absence of some nonperturbative contributions) has come from
duality arguments.  There was also a direct 2-loop computation 
which was incomplete; in this note
we explain the deficiency of the previous 2-loop calculation
and discuss the complete 2-loop computation in two different models. 
The corrected analysis yields a vanishing 2-loop contribution to
$\Lambda$ in these models.

\Date{October 1998}

\lref\verlong{E. Verlinde and H. Verlinde, ``Chiral Bosonization,
Determinants and the String Partition Function,'' Nucl. Phys.
{\bf B288} (1987) 357.} 
\lref\martinec{E. Martinec, ``Conformal Field Theory on a 
(Super)Riemann Surface,'' Nucl. Phys. {\bf B281} (1987) 157.} 
\lref\dHP{E. d'Hoker and D. Phong, ``The Geometry of String
Perturbation Theory,'' Rev. Mod. Phys. {\bf 60} (1988) 917.}
\lref\amsgl{J. Atick, G. Moore, and A. Sen, ``Some Global Issues
in String Perturbation Theory,'' Nucl. Phys. {\bf B308} (1988) 1.}
\lref\amscat{J. Atick, G. Moore and A. Sen, ``Catoptric Tadpoles,''
Nucl. Phys. {\bf B307} (1988) 221.}
\lref\rohm{R. Rohm, ``Spontaneous Supersymmetry Breaking in
Supersymmetric String Theories,'' Nucl. Phys. {\bf B237} (1984) 553.} 
\lref\fms{D. Friedan, E. Martinec and S. Shenker, ``Conformal Invariance,
Supersymmetry and String Theory,'' Nucl. Phys. {\bf B271} (1986) 93.}
\lref\verlinde{E. Verlinde and H. Verlinde, ``Mutliloop Calculations in
Covariant Superstring Theory,'' Phys. Lett. {\bf B192} (1987) 95.}
\lref\mor{A. Morozov, ``Two-Loop Statsum of Superstrings,'' Nucl. Phys.
{\bf B303} (1988) 343.}
\lref\morper{A. Morozov and A. Perelomov, ``Statistical Sums in 
Superstring Theory: Genus 2,'' Phys. Lett. {\bf B199} (1987) 209.} 
\lref\lfact{O. Lechtenfeld, ``Factorization and Modular Invariance of
Multiloop Superstring Amplitudes in the Unitary Gauge,'' Nucl. Phys.
{\bf B338} (1990) 403.} 
\lref\kks{S. Kachru, J. Kumar and E. Silverstein, ``Vacuum Energy
Cancellation in a Nonsupersymmetric String,'' hep-th/9807076.} 
\lref\ksU{S. Kachru and E. Silverstein, ``Self-Dual Nonsupersymmetric
Type II String Compactifications,'' hep-th/9808056.}
\lref\ksorb{S. Kachru and E. Silverstein, ``4d Conformal Field Theories
and Strings on Orbifolds,'' hep-th/9802183.}
\lref\adscft{J. Maldacena, ``The Large N Limit of Superconformal Field
Theories and Supergravity,'' hep-th/9711200\semi
S. Gubser, I. Klebanov and A. Polyakov, ``Gauge Theory Correlators from
Noncritical String Theory,'' Phys. Lett. {\bf B428} (1998) 105, 
hep-th/9802109\semi
E. Witten, ``Holography and Anti de Sitter Space,'' hep-th/9802150.}
\lref\lp{O. Lechtenfeld and A. Parkes, ``On Covariant Multi-loop
Superstring Amplitudes,'' Nucl. Phys. {\bf B332} (1990) 39.}

\lref\st{G. Shiu and S.H. Tye, ``Bose-Fermi Degeneracy and Duality
in Non-Supersymmetric Strings,'' hep-th/9808095.}
\lref\jeff{J. Harvey, ``String Duality and Nonsupersymmetric Strings,''
hep-th/9807213.}
\lref\ooguri{T. Eguchi and H. Ooguri, ``Chiral Bosonization on a
Riemann Surface,'' Phys. Lett. {\bf 187B} (1987) 127.}

\newsec{Introduction}

A new class of non-supersymmetric
type II models was recently introduced \refs{\kks,\ksU}.  These 
models have a simple mechanism for cancellation
of the 1-loop vacuum energy (due to a guaranteed bose-fermi
degeneracy).  A similar class of models was constructed in the
free-fermionic prescription in \st. 
In \kks\ a heuristic argument suggesting that
the cancellation should persist to all higher loops was presented.  
A broader motivation
arising from the correspondence between quantum-mechanical fixed lines
and flat dilaton potentials \ksorb\ in the AdS/CFT correspondence
\adscft\ also suggests that cancellation should occur.  Various
string duality computations now provide indirect evidence for
the absence of certain perturbative (and also certain nonperturbative)
contributions to $\Lambda$ in these models \refs{\jeff,\ksU,\st}.

In \kks\ a direct analysis of the 2-loop contribution
to the cosmological constant was also presented.  
It was argued that, in a particular gauge,
it cancels pointwise on the moduli space.  This was based on
an analysis of only the spin-structure-dependent part of the 
amplitude. 
This analysis was
incomplete.  In the gauge we chose, picture changing operators
coincide, and their operator product expansion produces poles.  
Some of these pole pieces cancel
between matter and ghost contributions, or constitute total derivatives
on the moduli space.  In general, however, there can be
finite pieces which arise from the
Taylor expansion of the spin-structure
dependent piece.\foot{We thank Greg Moore for
pointing out this error and for providing helpful constructive comments.}

In this note we rectify this situation for both the original model of 
\kks\ and for an even simpler model which also enjoys similar properties, 
and was introduced in \ksU.  This analysis will also appear 
in a revised version of \kks.  
In the simpler model, if we choose the gauge with coincident picture-changing
operators, summing over {\it all} spin structures
produces a high enough order zero in the spin-structure-dependent
piece to cancel against surviving poles in the spin-structure-independent
factor.  
We discuss this cancellation in \S4.   
In \S5, we analyze the Taylor expansion of the original set of
models \kks.  We show that after summing over spin structures
all potentially 
finite terms either cancel between different terms in the correlator
of picture changing operators or vanish directly.

Perturbative string calculations at higher genus are notoriously subtle.
In all of our analysis we assume that the prescription \lp\
(following e.g. \morper\ and \mor ) for the
supersymmetric 2-loop GSO phases is correct
(in particular, in \lfact\ it was argued that this prescription
leads to the correct phases upon factorization).  Adjusting 
the supersymmetric expression to incorporate the phases and twists of
our nonsupersymmetric 
models is relatively straightforward, and we see that the cancellation
persists if we make a further specialization of the gauge
choice.  Different gauge choices differ by total derivatives 
on the moduli space as explained in \refs{\verlinde,\amsgl}.  This
argument applies twist structure by twist structure in the
orbifold theory, as what it uses is deformation of a contour
integral of the BRST charge, which is invariant under the
orbifold group. The corresponding
boundary contributions cancel as discussed in \kks.
As explained in \kks, issues of modular invariance of
the gauge choice are simpler in our models due to shifts involved
in the orbifold group elements, which greatly reduce the modular
group in a given twist structure.  

\newsec{The Simpler Model}

In addition to the original model of \kks, we will also be considering
a new model introduced in \ksU.  Since the description of the new model
there is somewhat terse, we introduce it in greater detail here.
Like the model of \kks, it is an orbifold generated
by two elements $F$ and $G$.  We start from the type II string
compactified on a $T^6$ which is a product of six circles
at the self-dual radius $R = l_{s}/{\sqrt 2}$, where
$l_{s} = \sqrt{\alpha^\prime}$ is the string scale.  
$F$ and $G$ act on the left and right
moving degrees of freedom of the superstring as 

\bigskip
\vbox{\settabs 3 \columns
\+$S^1$&$F$&$G$\cr
\+1&$1$&$(s,-1)$\cr
\+2&$1$&$(s,-1)$\cr
\+3&$1$&$(s,-1)$\cr
\+4&$1$&$(s,-1)$\cr
\+5&$1$&$(0,s^2)$\cr
\+6&$(s,s)$&$1$\cr
\+&$(-1)^{F_R}$&$(-1)^{F_L}$\cr}    

\noindent
$s$ refers to a shift 
by $R/2=l_{s}/ 2 \sqrt{2}$.  As explained in \ksU, this
model has vanishing 1-loop vacuum energy (after integrating
over ${\rm Re}~ \tau$) due to the fact
that the trace of $G$ acting on level matched states 
in the $F$-twisted sector vanishes,\foot{The relevant fundamental 
domain for
the integral is the strip instead of the normal keyhole region due to
the shifts in $G$.} 
and
the other modular inequivalent contributions vanish by zero
modes due to effective supersymmetry.   

\newsec{2-loops with supersymmetry}

In this section we review the supersymmetric cancellation at
2 loops.  As explained for example in \refs{\dHP,\verlinde,\amsgl},
the type II string path integral can be written as
\eqn\path{
\int_{{\cal SM}_h}d\mu_0\int [dBdCdX]e^{-S}
\prod_{r=1}^{6h-6}(\eta_r,B)
\prod_{a=1}^{4h-4}\delta((\eta_a,B))
}
Here $B,C$ denote the $b,\beta$ and $c,\gamma$ ghosts, where
$(b,c)$ are the spin-(2,-1) conformal ghosts and $(\beta,\gamma)$
are the spin-(3/2,-1/2) superconformal ghosts.  $X$ denotes the
matter fields and $\eta_r$ and $\eta_a$ are Beltrami differentials
relating the metric and gravitino to the moduli and supermoduli.
In components,
\eqn\beltI{
(\eta_r,B)=\int\eta_{r\bar z}^z b_{zz}+\int\eta_{r\bar z}^+ \beta_{z+}
+\int\eta_{r z}^{\bar z} b_{\bar z\bar z}
+\int\eta_{r z}^- b_{\bar z -}
}
\eqn\beltII{
(\eta_a,B)=\int\eta_{a\bar z}^z b_{zz}+\int\eta_{a\bar z}^+ \beta_{z+}
+\int\eta_{a z}^{\bar z} b_{\bar z\bar z}
+\int\eta_{a z}^- b_{\bar z -}
}
As explained e.g. in \refs{\martinec,\dHP}, 
we can write the path integral measure on
supermoduli space in terms of
a fixed measure on 
moduli space
\eqn\detch{
d\mu_0=d\mu [sdet(\eta,\Phi)]^{-1}[sdet(\Phi,\Phi)]^{1/2}
}
Here $d\mu$ is a fixed measure on the supermoduli space
${\cal{SM}}_h$, integrated over a fixed domain independent
of the beltrami differentials.  $\Phi$ 
contains the $3h-3$ holomorphic and $3h-3$ antiholomorphic
2-differentials ($b$ ghost zero mode wavefunctions) and
the $2h-2$ holomorphic and $2h-2$ antiholomorphic 3/2-differentials
($\beta$ ghost zero modes).    

After choosing delta-function support for the worldsheet gravitinos,
and integrating out the supermoduli, one obtains a correlation
function of picture changing operators \verlinde 
\eqn\picop{
:e^\phi T_F: = c\partial\xi + {1\over 2} e^\phi\psi^\mu\partial X^\mu
-{1\over 4}\partial\eta e^{2\phi}b
-{1\over 4}\partial (\eta e^{2\phi} b)}
and other ghost insertions
\eqn\part{\eqalign{
\sum_{\alpha,\beta,twists}\int d\mu 
[sdet(\eta,\Phi)]^{-1}[sdet(\Phi,\Phi)]^{1/2}[dX][dB][dC]e^{-S}
(\hat\eta,b)^{6h-6}\xi (x_0)\cr
\prod_{a=1}^{2h-2}:e^\phi T_F(z_a):
\prod_{a=2h-3}^{4h-4}:e^{\bar\phi}\bar T_F(z_a):\bar\xi(\bar y_0)\cr}}
The superconformal ghosts $\beta=\partial\xi e^{-\phi}$,
$\gamma=\eta e^{\phi}$ are defined in terms of spin-0 and
spin -1 fermions $\xi,\eta$ and a scalar $\phi$ \fms. 
The spin-0 fermion $\xi$ has a zero mode on the surface which
is absorbed by the insertion of $\xi (x_0)$ in \part.
There is an anomaly in the ghost number $U(1)$ current which
requires insertions of operators with total ghost number
$2h-2$ to get a nonvanishing result. 
The correlation functions \part\ can be evaluated using the formulas
derived in e.g. \refs{\verlong,\ooguri}.

We will be interested in a gauge choice for which
ultimately $z_1,z_2\to\Delta_\gamma$, where
$\Delta_\gamma$ is a divisor corresponding to an
odd spin structure $\gamma$.    
We need to consider the (left-moving) spin-structure-dependent pieces
of the correlation function, the poles arising from
the spin-structure-independent local behavior of
the picture changing correlator, and the behavior of the
determinant \detch\ in this gauge. 
According to \verlinde\ we have the following contributions
to the spin-structure-dependent pieces of the 2-loop partition function.
The matter part of $T_F$ contributes 

\eqn\matt{
\sum_\delta \langle \delta|\gamma \rangle 
{{\theta[\delta]^4(0)\theta[\delta](z_1-z_2)}\over
{\theta[\delta](z_1+z_2-2\Delta_\gamma)}}}
Here $\delta\equiv (\alpha,\beta )$ encodes the
spin structure of the various contributions and
$\langle \delta|\gamma \rangle = e^{4\pi i(\alpha\gamma_2-\beta\gamma_1)}$
encodes the GSO phases \lfact.

Let us first, following \lp, take $z_1+z_2=2\Delta_\gamma$,
that is place $z_1+z_2$ at a divisor corresponding to the
canonical class,
without setting $z_1=z_2$.   
The contribution \matt\ then simplifies to
\eqn\nummatt{
\sum_\delta \langle \delta|\gamma \rangle \theta[\delta]^3(0)\theta[\delta](z_1-z_2)
=4\theta[\gamma]^4(z_1-\Delta_\gamma)
}
where in the last step we have used a Riemann identity.
The Riemann Vanishing Theorem then implies that
this vanishes identically as a function of $z_1$.
Thus in this case whatever poles arise as $z_1\to z_2$,
the identical zero from the spin structure sum cancels it.

Now turning to the ghost piece of the correlation function of
picture-changing operators, one obtains contributions
isomorphic to \nummatt\ as well as 
\eqn\ghost{
\omega_i(z_1){{\theta[\delta]^5(0)\partial_i\theta[\delta](2z_2-2\Delta_\gamma)}
\over{\theta^2[\delta](z_1+z_2-2\Delta_\gamma)}}
}
Here $\omega_i$ are the canonical basis of holomorphic
one-forms on the Riemann surface, satisfying
$\int_{a_i}\omega_j=\delta_{ij}$ and 
$\int_{b_i}\omega_j=\tau_{ij}$ where $\tau$ is the
period matrix for the surface.
Again simplifying this by first taking 
$z_1+z_2=2\Delta_\gamma$ we obtain
\eqn\numghost{
\sum_\delta \langle \delta|\gamma \rangle 
\partial_{z_1}(\theta[\delta]^3(0)\theta[\delta](z_1-z_2)) 
=4\partial_{z_1}(\theta[\gamma]^4(z_1-\Delta_\gamma))}
Because the right-hand side of this expression is a
derivative of 0 (by the Riemann vanishing theorem), it
vanishes identically.  Again any poles from the picture changing
OPEs are irrelevant \lp.

\newsec{2-loop cancellation in $F,G$ Model}

Let us consider the left-moving spin-structure-dependent pieces of
the 2-loop partition function in the model of section 2.
As explained in \kks, 
all supersymmetry-breaking twist structures
map by modular transformations to a canonical one in which there are twists 
$(1,1,F,G)$ in going around cycles $(a_1,a_2,b_1,b_2)$.  
The $F$ twist, which just involves an action of $(-1)^{F_R}$
and a symmetric shift, does not affect the $\theta$-function
characteristics and phases.  The effect of the $G$ twist
is via its $(-1)^{F_L}$ action.  This changes the $G$ eigenvalues
by a factor of $(-1)$ 
in the left-moving Ramond sector relative to a model
without the $(-1)^{F_L}$.  In our twist structure there is
a $G$ operator inserted on the second handle.  So the
effect of the $G$ twist is to change the phase of
terms in the spin structure sum with $\alpha_2=1/2$,
that is terms with Ramond sector states propagating around
handle 2.  This means that instead of  phases $\langle \delta|\gamma \rangle$  
we should have phases $\langle \delta|\gamma+(0,0,0,1/2) \rangle$.
We will set $\kappa\equiv\gamma+(0,0,0,1/2)$ to simplify the formulas.

Then after imposing $z_1+z_2=2\Delta_\gamma$, we
get for the
matter correlator

\eqn\nsnum{\eqalign{
\sum_\delta \langle \delta|\kappa\rangle 
\theta[\delta]^3(0)\theta[\delta](z_1-z_2)
\theta[\gamma](\sum_{i=1}^3w_i-3\Delta_\gamma)
{\prod_{i<j}E(w_i,w_j)\over{E(z_1,z_2)^2}}\cr
\times \biggl(2\pi p_i^\mu p_j^\mu\omega_i(z_1)\omega_j(z_2)+
10\partial_{z_1}\partial_{z_2}log E(z_1,z_2)
\biggr)\cr}}
where $E(z_1,z_2)$ is the prime form, which has the property
that it goes like $z_1-z_2$ as $z_1\to z_2$.
Here the $w_i,i=1,\dots,3h-3$ are the insertion points for
the $b$-ghosts.  In the full amplitude \verlinde, they
are contracted with beltrami differentials encoding the
gauge choice for the worldsheet metric.  
The spin-structure-dependent pieces here (i.e. those that depend
on the characteristic $\delta$) sum up to
\eqn\nssum{
4\theta[\kappa](z_1-\Delta_\gamma)^4
}
Because of the shift in the theta characteristic after the
spin structure sum, the right hand side no longer vanishes
identically.  However, if we choose  the reference spin
structure $\gamma$ 
such that both $\gamma$ and $\kappa\equiv\gamma+(0,0,0,1/2)$ are
odd, then we can impose $z_1,z_2\to\Delta_\gamma$ and
the right hand side of \nssum\ vanishes like
$(z_1-z_2)^4$ as $z_1\to z_2$.  
As $z_1\to z_2$ the determinant factor \detch\ produces another zero:
plugging in the delta function $\eta_a$ we obtain
\eqn\detzer{
[sdet(\eta,\Phi)]^{-1}\propto 
det(\eta_a,\Phi_b^{3/2})
=det(\Phi_b^{3/2}(z_a))
}
Here $\Phi_b^{3/2}, b=1,2$ form a basis of holomorphic 3/2-differentials. 
As the $z_a$ approach each other, the determinant \detzer\ goes
to zero.

Combining the spin structure sum with the determinant factor
we have a fifth order zero.
As we take the limit $z_1\to z_2$ we must take into account
the poles arising from the prime form $E(z_1,z_2)$ (which
goes like $z_1-z_2$ as $z_1\to z_2$).  
The $2\pi p_i^\mu p_j^\mu\omega_i(z_1)\omega_j(z_2)$
term in \nsnum has a second order pole, which is beaten by
our fifth order zero.  The other contributions yield
\eqn\mattleft{
det(\eta_a,\Phi_b^{3/2})
4\theta[\kappa]({1\over 2}(z_1-z_2))^4
\theta[\gamma](\sum w-3\Delta_\gamma)\prod_{i<j}E(w_i,w_j)
{{10}\over{(z_1-z_2)^4}}
}      
The poles arise from the correlator of picture changing operators.
The leading singularity in this correlator is proportional to 
${c\over{(z_1-z_2)^4}}$ where $c$ is the total central
charge (matter+ghost).  Since this vanishes, there should be
no such contribution after including the ghost terms in
the picture changing operators.  We will see this 
cancellation after
computing the ghost correlators.  In any case our fifth order zero
suffices to cancel this contribution even before including the
ghost piece, but after doing the spin structure sum.

The ghost correlator 
\eqn\ghostcorr{\eqalign{
\langle -{1\over 4} c\partial\xi(z_1) \left(2 \partial \eta e^{2\phi}b
+ \eta \partial e^{2\phi} b + \eta e^{2\phi}\partial b \right)(z_2)\rangle \cr
+
\langle -{1\over 4} \left(2 \partial \eta e^{2\phi} b + 
\eta \partial e^{2\phi} b + \eta e^{2\phi} \partial b \right)(z_1) c\partial
\xi (z_2)\rangle~.\cr}}   
gives rise to three types of contributions (see \verlinde\ eqn. (38-39) for the
complete expression).  One is
\eqn\ghostI{\eqalign{
& \sum_\delta \langle \kappa|\delta \rangle 
{{\theta[\delta](0)^5\theta[\delta](2z_2-2\Delta_\gamma)
\omega^i(z_1)\partial^i\theta[\gamma](z_1-z_2+\sum w -3\Delta_\gamma)}
\over{\theta[\delta](z_1+z_2-2\Delta_\gamma)^2}}
{{\prod_{i<j}E(w_i,w_j)}\over{E(z_1,z_2)^3}}\cr }
}
Setting $z_1+z_2=2\Delta_\gamma$ and summing over $\delta$ we
get for the spin-structure-dependent piece:
\eqn\ghostIsum{
4\theta[\kappa]({1\over 2}(z_1-z_2))^4\sim
(z_1-z_2)^4}
In this contribution, there is a third-order pole from
the prime forms, but a fifth-order zero \ghostIsum\ from
the spin structure sum times the determinant factor
\detzer, so the contribution cancels as
$z_1\to z_2$.  

The second ghost contribution takes the form
\eqn\ghostII{\eqalign{
& \sum_\delta \langle \kappa|\delta \rangle 
{{\theta[\delta](0)^5\theta[\delta](2z_2-2\Delta_\gamma)
\theta[\gamma](z_1-z_2+\sum w -3\Delta_\gamma)}
\over{\theta[\delta](z_1+z_2-2\Delta_\gamma)^2}}\cr
& \times {{\prod_{i<j}E(w_i,w_j)}\over{E(z_1,z_2)^3}}
\partial_{z_1}log({{\prod_jE(z_1,w_i)}\over{E(z_1,z_2)^5\sigma(z_1)}})
+(z_1\leftrightarrow z_2)\cr}
}
where $\sigma(z_1)$ is defined in \verlinde\ and has no zeroes
or poles.
Taking $z_1+z_2=2\Delta_\gamma$ and summing over $\delta$ we get
\eqn\ghostIIsum{
4\theta[\kappa]({1\over 2}(z_1-z_2))^4
{{\prod_{i<j}E(w_i,w_j)}\over{E(z_1,z_2)^3}}
\partial_{z_1}log({{\prod_jE(z_1,w_i)}\over{E(z_1,z_2)^5\sigma(z_1)}})
+(z_1\leftrightarrow z_2)}
This contribution has again a fifth order zero from the
spin-structure sum times the determinant
(as in \ghostIsum ).  
There is also a
fourth-order pole from
the prime forms.  The contribution is
\eqn\ghostIIfin{\eqalign{
& det(\Phi_b^{3/2}(z_a))
4\theta[\kappa]({1\over 2}(z_1-z_2))^4
\prod_{i<j}E(w_i,w_j)
{{-5}\over{(z_1-z_2)^4}}
\theta[\gamma](\sum w-3\Delta_\gamma)\cr
&+(z_1\leftrightarrow z_2)\cr }
}
This piece cancels against the matter contribution
\mattleft.  As mentioned above this encodes the
fact that the total central charge vanishes.
Again taking into account \detzer\ we in fact obtain
cancellation directly.

Now let us consider the remaining ghost contributions.
These arise from the correlator
\eqn\lastghform{
\langle c\partial\xi(z_1)~(\partial e^{2\phi})\eta b (z_2) \rangle 
+
\langle (\partial e^{2\phi})\eta b (z_1)~c\partial\xi(z_2)\rangle 
}
They take the form
\eqn\ghostIII{\eqalign{
& det(\Phi_b^{3/2}(z_a))
\sum_\delta \langle \kappa|\delta \rangle 
{{\theta[\delta](0)^5
\omega_i(z_1)\partial^i\theta[\delta](2z_2-2\Delta_\gamma)
\theta[\gamma](z_1-z_2+\sum w -3\Delta_\gamma)}
\over{\theta[\delta](z_1+z_2-2\Delta_\gamma)^2}}
{{\prod_{i<j}E(w_i,w_j)}\over{E(z_1,z_2)^3}}\cr
&+(z_1\leftrightarrow z_2)\cr }
}
In the first term of this contribution \lastghform, 
setting $z_1+z_2=2\Delta_\gamma$ and
summing on $\delta$ gives
\eqn\ghostIIsum{
\omega_i(\Delta_\gamma)
\partial^i\theta[\kappa](z_1-z_2)^4{1\over{(z_1-z_2)^3}}
}
plus other terms which have lower-order poles.
The spin structure sum only gives a third order zero here due
to the derivative acting on the theta functions.  The
$ det(\Phi_b^{3/2}(z_a))$ factor gives an additional zero
which kills this contribution as $z_1\to z_2$.

Thus we see that in the $(F,G)$ model we can pick a gauge for
the worldsheet metric as well as gravitino, for which 
all the contributions cancel,
after summing over all spin structures and using appropriate Riemann
identities.  Note that the cancellations explained above occur
via the spin structure sum of each term in the correlator
of picture-changing operators separately, so that the shifts twisting
some of the bosons do not affect the argument.
The above cancellation alone is not enough to guarantee that the
two-loop amplitude cancels; one needs the cancellation of the boundary
contributions that potentially arise due to the ambiguity in
the choice of gauge slice for the gravitini \amscat.  In the $(F,G)$
model these appear to cancel similarly to the analyis
in the original models \kks.    
Note that although the bulk cancellations we exhibit here would hold
also for example for the type of model studied in \rohm\ 
(generated by a translation combined with $(-1)^{F_L+F_R}$), 
those models have a nonzero one-loop vacuum energy and non-zero
one-loop tadpoles (as well as tachyons, for some range of the moduli VEVs).  
Therefore, the
boundary contributions in the latter case would destroy the cancellation.
Nonetheless it is somewhat surprising that the bulk contribution can
be made to cancel by choosing an appropriate gauge in that model.

\newsec{Taylor Expansion in Original Models}

In this section we will present the analysis of the coincident
gauge for the original models of \kks\ (see that paper for
the definition of the model).  
As in \kks, we only need to analyze the twist structure
$(1,1,f,g)$.  The $f$ twist affects the characteristics of 
some of the $\theta$ functions (arising from twisted fields)
by shifting them by $(0,0,1/2,0)$ -- we shall denote this
as a shift by ${1\over 2}L$.  $\kappa$ is, as in \S4,
$\gamma + (0,0,0,1/2)$, and we choose $\gamma$ such
that both $\gamma$ and $\kappa$
are odd. 

The correlation function of
the matter part of the picture changing operators breaks into
two contributions.  The terms involving  
$\langle \psi^i\partial X^i(z_1)\psi^i\partial X^i(z_2) \rangle$ with
$i=5,\cdots 10$ give 
\eqn\mattl{\eqalign{
\sum_\delta \langle \kappa|\delta \rangle 
{{\theta[\delta](0)^2\theta[\delta+{1\over 2}L](0)^2\theta[\delta](z_1-z_2)}
\over{\theta[\delta](z_1+z_2-2\Delta_\gamma)}}\cr
\times ( p_i^\mu\omega^i(z_1)p_j^\mu\omega^j(z_2)
{1\over{E(z_1,z_2)^2}}
+{6\over{E(z_1,z_2)^2}}\partial_{z_1}\partial_{z_2}log E(z_1,z_2))\cr 
\times det(\Phi^{3/2}_a(z_b))}}

Upon setting $z_1+z_2=2\Delta_\gamma$ we can cancel the denominator
against one factor in the numerator to get 
\eqn\mattsimp{
\sum_\delta \langle \kappa |\delta \rangle 
\theta[\delta](0)\theta[\delta](z_1-z_2)
\theta[\delta+{1\over 2}L](0)^2
=4\theta[\kappa]({1\over 2}(z_1-z_2))
\theta[\kappa+{1\over 2}L]({1\over 2}(z_1-z_2)) 
}
for the spin-dependent piece of this correlator. 
Because $\kappa$ is an odd spin structure, this vanishes like
$(z_1 - z_2)^{2}$.  The factor of $det(\Phi^{3/2}_a(z_b))$ in
\mattl\ also vanishes like $(z_1-z_2)$ as $z_1 \to z_2$, so
all in all \mattl\ has a $(z_1-z_2)^3$ multiplying the
prime forms.  However, since the prime-forms are yielding poles
as $z_1 \to z_2$, it remains to check that there are no finite
pieces in \mattl.

The terms proportional to ${1\over E(z_1,z_2)^2}$ times the loop
momenta clearly vanish in
the limit, since there is only a second order pole from the prime forms
which cannot cancel the third order zero we found from the spin structure
sum and the superdeterminant.  This leaves the term which goes like
${1\over E(z_1,z_2)^2} \partial_{z_1}\partial_{z_2}log E(z_1,z_2)$.
Using the fact that $E(z_1,z_2)$ has a Taylor expansion of the form
\eqn\eexp{E(z_1,z_2) \sim \sum_{n=0}^{\infty} c_n (z_1 - z_2)^{2n+1}}
as $z_1 \to z_2$, one sees that this combination of prime forms has
an expansion
\eqn\pexp{{1\over E(z_1,z_2)^2} \partial_{z_1} \partial_{z_2} log E(z_1,z_2)
\sim {\sum_{n=-2}^{\infty} d_{n} (z_1-z_2)^{2n}}}
On the other hand, the determinant factor is an $\it odd$ function of
$z_1 - z_2$ with an expansion of the form
\eqn\detexp{det(\Phi^{3/2}_a(z_b)) \sim \sum_{m=0}^{\infty} e_m (z_1-z_2)^{2m+1}}
while the sum over spin structures \mattsimp\ is an even function with a
second order zero at $z_1=z_2$.  From these facts, it is easy to see that
the full expression \mattl\ has an expansion of the form
\eqn\mattexp{\sum_{j=0}^{\infty} f_{j} (z_1-z_2)^{2j-1}}
as $z_1 \to z_2$.  

Examining \mattexp, we see that

\noindent
$\bullet$ There are no finite contributions as $z_1 \to z_2$.

\noindent
$\bullet$ There is a (gauge artifact) pole as $z_1 \to z_2$; in fact 
this pole receives contributions from the various matter and ghost correlators
proportional to the matter/ghost central charges, and hence cancels once
all of the terms are taken into account (since $c_{tot} = 
c_{matter} + c_{ghost}=0$).  We will see this explicitly once we
compute the remaining matter and ghost contributions.

The second type of matter correlator arises from contracting
the $\psi^{i}\partial X^{i}(z_1) \psi^{i}\partial X^i (z_2)$ with 
$i=1,\cdots 4$.   
This leads to a contribution
\eqn\mattwo{\eqalign{\sum_{\delta}\langle \kappa | \delta \rangle {{\theta[\delta](0)^3
\theta[\delta + {1\over 2}L](0) \theta[\delta + {1\over 2}L] (z_1-z_2)}\over
{\theta[\delta](z_1 + z_2 - 2\Delta_{\gamma})}}\cr
\times (p_i^\mu \omega^i (z_1) p_j^\mu \omega^j (z_2) {1\over E(z_1,z_2)^2} +
{4\over E(z_1,z_2)^2}\partial_{z_1}\partial_{z_2} log E(z_1,z_2))\cr
\times det(\Phi^{3/2}_a(z_b))}} 

Choosing $z_1 + z_2 = 2\Delta_{\gamma}$, the spin sum in \mattwo\
simplifies to
\eqn\spintwo{\sum_{\delta} \langle \kappa | \delta\rangle \theta[\delta](0)^2
\theta[\delta + {1\over 2}L](0) \theta[\delta + {1\over 2}L] (z_1-z_2)}
which, after applying a Riemann identity, becomes
\eqn\spintwon{4 \theta[\kappa]({1\over 2}(z_1-z_2))^{2} 
\theta[\kappa+{1\over 2}L]
({1\over 2} (z_1-z_2))^2}
So in fact after summing over spin structures this looks the same as the 
spin sum of the first type of matter contribution \mattsimp.
Again, it vanishes like $(z_1-z_2)^2$ as $z_1 \to z_2$.

Now, the argument for the cancellation proceeds as it did for the first
type of matter contribution.
The terms involving only the ${1\over E(z_1,z_2)^2}$ multiplying loop
momenta only have a second order pole, which cannot cancel the third
order zero coming from the determinant times the spin structure sum \spintwon. 
The terms involving higher inverse powers of the prime forms lead to 
a simple pole (which cancels after summing over matter and ghosts, as it is
proportional to the total central charge) and no finite contributions.

Next, let us consider the terms in the correlator of picture changing
operators coming from the ghost part of the worldsheet supercurrent
\ghostcorr.  There are three types of terms that arise \verlinde.  
We are in the twist structure $(1,1,f,g)$.  As in the matter
sector, the $f$ twist affects the characteristics of the $\theta$-functions
arising in the worldsheet correlation functions and determinants.
We will denote the shift in the characteristic, which is
$(0,0,1/2,0)$, as ${1\over 2}L$.   
The
first type of contribution is
\eqn\firstgh{\eqalign{
\sum_\delta \langle \kappa |\delta \rangle 
{{\theta[\delta](0)^3\theta[\delta+{1\over 2}L](0)^2
\theta[\delta](2z_2-2\Delta_\gamma)\theta(z_1-z_2+\sum w-3\Delta)}
\over{\theta[\delta](z_1+z_2-2\Delta_\gamma)^2 E(z_1,z_2)^3}}\cr
\times det (\Phi_a(z_b)) {{\prod E(z_1,w)}\over{\prod E(z_2,w)}}
\partial_{z_1}log({{\prod E(z_1,w)}\over{E(z_1,z_2)^5\sigma(z_1)}})
\cr
+(z_1\leftrightarrow z_2)\cr}}
The second is
\eqn\secondgh{
\eqalign{
\sum_\delta \langle \kappa |\delta \rangle 
{{\theta[\delta](0)^3\theta[\delta+{1\over 2}L](0)^2
\omega_i(z_1)\partial^i\theta[\delta](2z_2-2\Delta_\gamma)
\theta(z_1-z_2+\sum w-3\Delta)}
\over{\theta[\delta](z_1+z_2-2\Delta_\gamma)^2 E(z_1,z_2)^3}}\cr
\times det (\Phi_a(z_b)) {{\prod E(z_1,w)}\over{\prod E(z_2,w)}}
\cr
+(z_1\leftrightarrow z_2)\cr
}}
The third is
\eqn\thirdgh{
\eqalign{
\sum_\delta \langle \kappa |\delta \rangle 
{{\theta[\delta](0)^3\theta[\delta+{1\over 2}L](0)^2
\theta[\delta](2z_2-2\Delta_\gamma)
\omega_i(z_1)\partial^i\theta(z_1-z_2+\sum w-3\Delta)}
\over{\theta[\delta](z_1+z_2-2\Delta_\gamma)^2 E(z_1,z_2)^3}}\cr
\times det \Phi_a(z_b) {{\prod E(z_1,w)}\over{\prod E(z_2,w)}}
\cr
+(z_1\leftrightarrow z_2)\cr
}}

Setting $z_1+z_2=2\Delta_\gamma$ and doing the spin structure
sum we find for the spin-structure-dependent pieces of 
contributions \firstgh\ and \thirdgh :
\eqn\spsumI{\eqalign{
\sum_\delta \langle \kappa |\delta \rangle 
\theta[\delta](0)\theta[\delta+{1\over 2}L](0)^2
\theta[\delta](2z_2-2\Delta_\gamma)\cr
=\theta[\kappa](z_2-\Delta_\gamma)^2
\theta[\kappa+{1\over 2}L](z_2-\Delta_\gamma)^2\cr
\sim (z_1-z_2)^2+c_4(z_1-z_2)^4+\dots\cr}}
for some constant $c_4$
where in the last line we expanded the result in
a Taylor expansion around $z_1=z_2$.
For contribution \secondgh\ we get
\eqn\spsumII{\eqalign{
\sum_\delta \langle \kappa |\delta \rangle  
\theta[\delta](0)\theta[\delta+{1\over 2}L](0)^2
\omega_i(z_1)\partial^i\theta[\delta](2z_2-2\Delta_\gamma)\cr
=
\partial_{z_1}\biggl( \theta[\kappa]({1\over 2}(z_1-z_2))^2
\theta[\kappa+{1\over 2}L]({1\over 2}(z_1-z_2))^2\biggr)\cr
\sim (z_1-z_2)+b_3(z_1-z_2)^3+\dots\cr }}

As for the matter contributions, although the spin structure
sums give vanishing contributions, they multiply singularities
arising from the prime forms $E(z_1,z_2)$ and we must analyze
the potential finite terms in the Taylor expansion.
Let us consider first \firstgh.  There are two types of 
contributions here.  After doing the spin structure
sum as above the first takes the form
\eqn\firstform{\eqalign{
-5{{\partial_1 E(z_1,z_2)}\over{E(z_1,z_2)^4}}[z_{12}^2+c_4z_{12}^4+\dots]
[z_{12}+e_3z_{12}^3+\dots]{{\prod E(z_1,w)}\over{\prod E(z_2,w)}}
+(z_1\leftrightarrow z_2)\cr}}
where we denote $z_1 - z_2$ by $z_{12}$.
Here the second factor comes from the spin structure sum, the
third from the Taylor expansion of the determinant about 
$z_1=z_2$ (where $e_3$ is some constant).  
We should emphasize what is meant here by $(z_1\leftrightarrow z_2)$.
We are computing a correlation function of picture changing
operators.  The ghost piece of this correlator has the form
\ghostcorr.  So for example the second term in \ghostcorr\ corresponds to the
term denoted $z_1\leftrightarrow z_2$ in \firstform.  So
in particular the second term involves interchanging the operators
in the ghost correlator, without changing $z_1$ to $z_2$ in the
determinant factor.  
The first and fourth
factors involving the prime forms encode the physical poles
and zeroes of the correlator.
The leading singularity from the prime forms here comes from
the $1/z_{12}^4$ term in the expansion of the prime form factors.
Therefore only the leading term in the Taylor expansion of the
spin structure sum and determinant factors potentially survive
(so we can ignore the terms proportional to $c_4$ or $e_3$,
which give fifth-order zeroes).
Similarly expanding the prime forms $E(z_1,z_2)$ gives a subleading
term with only a $1/z_{12}^2$ pole, which is cancelled by the third
order zero coming from the leading piece of the spin structure
sum times determinant.  

Putting the factors together, we see that the leading piece is 
a simple pole in $z_{12}$.  
The first three factors in \firstform\
are the same in the term with $z_1\leftrightarrow z_2$.   
When we include the term with $z_1\leftrightarrow z_2$, they
multiply the 
prime form factor 
${{\prod E(z_1,w)}\over{\prod E(z_2,w)}}+
{{\prod E(z_2,w)}\over{\prod E(z_1,w)}}$.  This is
even under $z_1\leftrightarrow z_2$.  In our Taylor expansion
it therefore becomes of the form ${\cal O}(1)+f_2z_{12}^2+\dots$,
and only the first term contributes.  
Therefore in Taylor expanding the contribution \firstform, we
get a pole piece 
plus higher order terms which vanish in the
limit $z_1\to z_2$.  In particular, no finite pieces survive.
What is the interpretation of the pole piece?  It is proportional to
the ghost central charge, and precisely cancels the pole piece 
coming from the matter contribution.

The second type of contribution in \firstgh\ takes the form
\eqn\firstformb{\eqalign{
{1\over{E(z_1,z_2)^3}}[z_{12}^2+\dots][z_{12}+\dots]
{{\prod E(z_1,w)}\over{\prod E(z_2,w)}}\partial_1 log({{\prod E(z_1,w)}
\over{\sigma(z_1)}})
\cr}}
where the $\dots$ denotes terms which vanish automatically as
$z_1\to z_2$.  The leading pole from the prime forms here is
cubic.  Before including the $z_1\leftrightarrow z_2$ term there
is a finite piece obtained by multiplying this times the third
order zero obtained from the spin structure sum and determinant
factors.  The spin structure sum is even under the interchange
of $z_1$ and $z_2$ in this case, and as discussed above the
determinant factor is the same in both terms.  The factor
${1\over{E(z_1,z_2)^3}}$ does change sign between the two
terms, however.  So when we add the $(z_1\leftrightarrow z_2)$ term
the contribution cancels.

Let us now consider the contribution \secondgh.  This gives a
contribution of the form
\eqn\secondform{
{1\over{E(z_1,z_2)^3}}[z_{12}+\dots][z_{12}+\dots]
\biggl({{\prod E(z_1,w)}\over{\prod E(z_2,w)}}+
{{\prod E(z_2,w)}\over{\prod E(z_1,w)}}\biggr)
}
Here similarly to the above analysis we took into account the
relative sign of the two contributions in \ghostcorr\ and included
the $z_1\leftrightarrow z_2$ contribution.  The last factor
here is even under interchange of $z_1$ and $z_2$, so its
Taylor expansion is of the form $1+h_2z_{12}^2+\dots$ for
some constant $h_2$.  The leading contribution here is
a simple pole, and 
there is no finite contribution.

Unlike the previous simple poles we have encountered, the pole
encountered here does not cancel with the other matter and ghost
contributions (it is $\it not$ one of the pieces which would have
contributed to the ${c\over z^{4}}$ pole in the OPE of
picture changing operators before accounting for
spin structure sums and determinant factors).   
However, on general grounds we expect such gauge artifact poles to
constitute total derivatives on moduli space.  
Otherwise the invariance of the path integral on gauge slice
would be lost.  In this case, we can
argue for that conclusion as follows.  The pole we are discussing 
receives a ${1\over {(z_1-z_2)^{3}}}$ contribution from the prime forms
which is softened to ${1\over {(z_1-z_2)^{2}}}$ by the theta function
zero (and then to a simple pole by the determinant factor).  
In the OPE of picture changing operators, the 
${1\over {(z_1-z_2)^2}}$ divergence is multiplied by the stress-energy
tensor, which gives a derivative with respect to the metric and therefore
the moduli.  The term we
are finding is part of this total derivative.  In the gauge we have
chosen, it is the only non-vanishing piece (the other pieces vanish
even before integration over the moduli space).  However, since there
cannot be gauge artifact poles, we expect it to integrate to zero (which one
can argue for using the boundary analysis in \kks).

Finally let us consider the last ghost contribution \thirdgh.
This contribution takes the form
\eqn\thirdform{\eqalign{
{1\over{E(z_1,z_2)^3}}[z_{12}^2+\dots][z_{12}+\dots]
{{\prod E(z_1,w)}\over{\prod E(z_2,w)}}\cr
+(z_1\leftrightarrow z_2)\cr
}}
In this contribution before including the
$z_1\leftrightarrow z_2$ contribution there is a 
potential finite term from the third order pole
multiplying a third order zero in $z_{12}$.  
Here again, in the limit $z_1\to z_2$ 
every factor except the first is the same in
the two terms.  The first factor ${1\over{E(z_1,z_2)^3}}$ has
the opposite sign in the two terms.  Thus again after including
the $z_1\leftrightarrow z_2$ term the contribution cancels.

\newsec{Discussion}

In the original models of the type described in \kks,
as well as in the new model described in \ksU, 
we see here that the two loop contribution cancels in the
gauge with coincident picture changing operators
after properly taking into account the 
singularities.  

This analysis, combined with the analysis of boundary
contributions presented in \kks, yields a vanishing two
loop cosmological constant in these models.  Note that
we computed this contribution in a particular
twist structure, in the gauge where $z_1$ and $z_2$ approach 
a particular branch point $\Delta_\gamma$.  Performing
a modular transformation moves us to a different twist structure--as
explained in \kks\ one can recover all twist structures in
this way--but also permutes the branch points and in general
will change the gauge choice.\foot{We thank Z. Kakushadze for
discussions on this point.}  However at a given twist structure,
one choice of insertion points for picture changing operators
is related to another by total derivative terms which cancel
in our model \kks.  
Therefore in a fixed gauge we are in fact not finding cancellation
pointwise on the moduli space, but instead only after integration
(using the formal argument of \verlinde\ to reduce the integrals to
boundary contributions).  A modular invariant integrand is obtained
on the other hand if we let the gauge choice get modular tranformed
along with the
twist structure.  Needless to say it would be interesting
to compute the integral directly in some fixed gauge to check the
cancellation independently.

\centerline{\bf{Acknowledgements}}

\noindent
We would like to thank G. Moore for pointing out the deficiency in
the 2-loop computation in \kks, and for extremely helpful discussions.
We are also grateful to E. d'Hoker, J. Harvey, Z. Kakushadze,
J. Kumar, S. Shenker, H. Verlinde and E. Witten
for helpful discussions. 
This work was initiated at the Amsterdam Summer Workshop on String Theory
and continued while E.S. was enjoying the hospitality of the Aspen
Center for Physics.
S.K. is supported
by NSF grant PHY-94-07194, by DOE contract DE-AC03-76SF0098, by an
A.P. Sloan Foundation Fellowship and by a DOE OJI Award.  E.S. is supported
by DOE contract DE-AC03-76SF00515.

\listrefs
\end